\documentclass[12pt]{JHEP3}
\usepackage{epsfig}
\usepackage{amssymb}
\usepackage{bm}
\usepackage{amsmath}
\newcommand{\p}{\partial}

\makeatletter
\@addtoreset{equation}{section}
\makeatother

\title{
Three-Body Nuclear Forces \\
from a Matrix Model
}
\author{
Koji Hashimoto$^{*,a}$ and 
Norihiro Iizuka$^{\dagger,b}$\\
${^*}$ {\it Mathematical Physics Lab., RIKEN Nishina Center,
Saitama 351-0198, Japan}\\
${}^\dagger$
{\it Theory Division, CERN, CH-1211 Geneva 23, Switzerland}\\
$^a$ E-mail: \email{koji(at)riken.jp}\\ 
$^b$ E-mail: \email{norihiro.iizuka(at)cern.ch}\\
}

\abstract{
We compute three-body nuclear forces at short distances 
by using the nuclear matrix model of holographic QCD proposed in our
previous paper with P.~Yi.  
We find that the three-body forces at short distances
are repulsive 
for 
(a) aligned three neutrons with averaged spins,
and (b) aligned proton-proton-neutron / proton-neutron-neutron. 
These indicate that 
in dense states of neutrons such as 
cores of neutron stars, or in Helium-3 / tritium nucleus, 
the repulsive forces are larger than 
the ones estimated from two-body forces only. }

\preprint{
{\normalsize CERN-PH-TH-2010-063} \\
{\normalsize RIKEN-MP-2}\\
}

\begin{document}
\setcounter{page}{1}


\noindent
\section{Introduction.} 
\label{sec1}

{
One of the fundamental ingredients of nuclear physics is the nuclear force
with which point-like nucleons interact with each other. A variety of
aspects of nuclear forces results in the protean metamorphosis of nuclei,
the bound states of nucleons. 
It is known that in nuclear forces 
there are forces that  
can not be explained by two-body forces only, one of which is the three-body force. 
The three-body forces play important role, for example,  in   
reproducing excitation spectra of light nuclei, 
or explaining equations of states for high-density baryon matters 
such as supernovae and neutron stars.
However, in spite of the
long history of nuclear physics, the bulk properties of three-body nuclear
forces are yet to be revealed. 

The main obstacle for revealing the various aspects of nuclear forces is obvious: QCD is
strongly coupled and thus difficult to solve. 
In this paper, 
by using a nuclear matrix model of holographic QCD 
which we have derived together with P.~Yi in \cite{Hashimoto:2010je}, we 
explicitly compute a three-body nuclear force in a large $N_c$ holographic QCD. 
The two-body nuclear force was already computed in \cite{Hashimoto:2010je}.

For the derivation of our matrix model \cite{Hashimoto:2010je} 
we use the gauge/string duality (the AdS/CFT
correspondence) \cite{Maldacena:1997re,Gubser:1998bc,Witten:1998qj}
applied to a D4-D8 system \cite{Sakai:2004cn,Sakai:2005yt} 
of a large $N_c$ QCD at a large 'tHooft coupling $\lambda$. 
Precisely speaking, our matrix model is a low-energy
effective field theory on baryon vertex D4-branes 
\cite{Witten:1998xy} in the D4-D8
holographic model \cite{Sakai:2004cn,Sakai:2005yt} of large $N_c$ QCD. 
The matrix model describes
$k$-body baryon systems with arbitrary $k$, where
the size of the matrix is given by this $k$, based on the fact that 
baryons are wrapped D-branes on sphere \cite{Witten:1998xy} 
in the gravity side of the gauge/string duality. In the previous work
\cite{Hashimoto:2010je}, in addition to the derivation of the matrix
model, the cases with $k=1$ (baryon spectrum) and $k=2$
(two-body nuclear force) were studied. For $k=2$, it was found that 
a universal repulsive core exists for any baryon states with two flavors. 
Since our matrix model is not a phenomenological model for multi-baryon
systems, but based on a firm ground of the gauge/string duality in string theory, 
it is natural to
extend the analysis of our matrix model to derive the three-body 
nuclear forces. In this paper, 
we continue the analysis to the $k=3$ case, {\it i.e.}~we 
study the short-range three-body
nuclear force, using the matrix model.

Although generic configurations of three baryons can be treated in the
matrix model, as the computations are involved 
and thus not so illuminating,
in this paper we shall concentrate on 
two particular examples: (a) three neutrons 
with spins averaged and (b) proton-proton-neutron (and
proton-neutron-neutron),  
both aligned on a line with equal spacings. 
System with spin averaged is rather typical for dense states of 
multi-baryons such as cores of neutron stars.
The latter is related to Helium-3 and tritium nucleus. For both cases,  
the resultant three-body potential 
is positive, {\it i.e.} repulsive. 
It scales as $N_c/\lambda^2 r^4$ (where $r$
is the inter-nucleon distance), in contrast to the two-body repulsive
core $\sim N_c/\lambda r^2$. 
As the region of validity is
at short range, $1/\sqrt{\lambda}M_{\rm KK}\ll r \ll 1/M_{\rm KK}$
(where $M_{\rm KK}\sim {\cal O}(1$ GeV)), the three-body potential is
suppressed compared to the two-body potential by $\sim 1/\lambda (r M_{KK})^2 \ll 1$. 
However at very short distances, {\it i.e.} at high dense states of nucleons, three-body 
forces are not small.

The organization of this paper is as
follows. We first review the matrix model and the 
two-body calculation shown in 
\cite{Hashimoto:2010je}. 
Then in section 3, we calculate the three-body 
forces. First, as an exercise, we treat the case with 
spin/isospins aligned classically,
and find that the three-body force vanishes for this case, which is
consistent with the soliton approach \cite{Hashimoto:2009as}. 
After that, we 
proceed to generic three-body forces with quantum spin/isospins. 
The final section is devoted for discussions.


\vspace{2mm}
\noindent
\section{Review: a matrix model and two-body interactions.} 
\label{sec2}

The procedures of the computation of the three-body nuclear forces is
quite analogous to the two-body case performed in
\cite{Hashimoto:2010je}. Here we provide a summary of the matrix model
action and the computation of the two-body nuclear forces of
\cite{Hashimoto:2010je}. 

\subsection{A matrix model action.}

In \cite{Hashimoto:2010je}, we proposed with P.~Yi a $U(k)$ matrix model which
describes generic $k$-body interaction of nucleons. Note that the rank of 
gauge group $U(k)$ is not at all related to the number of colors $N_c$ but just the number of nucleons $k$. The matrix model
action is quite simple, 
\begin{eqnarray}
S &=& 
\frac{\lambda N_c M_{\rm KK}}{54 \pi} 
\int\! dt \; {\rm tr}_k 
\left[
(D_0 X^M)^2 -\frac23 M_{\rm KK}^2 (X^4)^2
+ D_0 \bar{w}^{\dot{\alpha}}_i D_0 w_{\dot{\alpha}i} 
- \frac16  M_{\rm KK}^2 \bar{w}^{\dot{\alpha}}_i w_{\dot{\alpha}i}
\right.
\nonumber \\
&&\left.
+ \frac{3^6 \pi^2}{4 \lambda^2 M_{\rm KK}^4} 
\left(\vec{D}\right)^2
+  \vec{D}\cdot \vec{\tau}^{\;\dot{\alpha}}_{\;\;\;\dot{\beta}}
\bar{X}^{\dot{\beta}\alpha} X_{\alpha \dot{\alpha}}
+  \vec{D}\cdot \vec{\tau}^{\;\dot{\alpha}}_{\;\;\;\dot{\beta}}
\bar{w}^{\dot{\beta}}_{i} w_{\dot{\alpha}i}
\right]
+  N_c \int \! dt\; {\rm tr}_k A_0 \, .
\label{mm}
\end{eqnarray}
The peculiar property of this matrix model is the simplicity: changing
the number of the nucleons $k$ is available just by choosing $U(k)$ for
the gauge group of the matrix model. In \cite{Hashimoto:2010je} it was
demonstrated how to compute the baryon spectrum $(k=1)$ and two-body
nuclear forces $(k=2)$ at short distances. 

To be concise, here we briefly describe the matter content of the matrix
model (\ref{mm}). The model has a unique scale 
$M_{\rm KK}$, and $\lambda = N_c g_{\rm QCD}^2$ 
is the 'tHooft coupling constant of QCD, with the number of colors
$N_c$. 
The field content is summarized in the following
table.   
\begin{center}
\begin{tabular}{|c|c||c|c|c|}
\hline
 field & index & $U(k)$ & $SU(N_f)$ & $SU(2)\times SU(2)$\\\hline
\hline
$X^M(t)$ & $M=1,2,3,4$ & adj. & ${\bf 1}$& $({\bf 2}, {\bf 2})$\\\hline
$w_{\dot{\alpha}i}(t)$ & $\dot{\alpha}=1,2$; $i=1,\cdots,N_f$
& ${\bf k}$ &${\bf N_f}$ & $({\bf 1},{\bf 2})$  \\\hline
$A_0(t)$ & & adj. & ${\bf 1}$ & $({\bf 1}, {\bf 1})$ \\\hline
$D_s(t)$& $s=1,2,3$ & adj. & ${\bf 1}$& $({\bf 1},{\bf 3})$\\\hline
\end{tabular}
\end{center}
The dynamical fields are only $X^M$ and $w_{\dot{\alpha}i}$, while $A_0$
and $D_s$ are auxiliary fields. In writing these fields, the indices for
the gauge group $U(k)$ are implicit. In this paper we consider only the
two-flavor case $N_f=2$ for simplicity.
The symmetry of this matrix quantum mechanics is
$U(k)_{\rm local}\times SU(N_f)\times SO(3)$ 
where the last factor $SO(3)$ is the spatial rotation, which, together
with a holographic dimension, forms a broken
$SO(4)\simeq SU(2)\times SU(2)$ shown in the table. The breaking is due
to the mass terms for $X^4$ and $w_{\dot{\alpha}i}$.  
In the action, the trace is over these $U(k)$ indices, and 
the definition of the covariant derivatives is
$D_0 X^M \equiv \partial_0 X^M -i[A_0, X^M]$,
$D_0 w \equiv \partial_0 w -i w A_0$ and 
$D_0 \bar w \equiv \partial_0 \bar w + i A_0 \bar w$.   
The spinor indices of $X$ are defined as
$X_{\alpha\dot{\alpha}}\equiv X^M(\sigma_M)_{\alpha\dot{\alpha}}$ and
$\bar{X}^{\dot{\alpha}\alpha}
\equiv X^M(\bar{\sigma}_M)^{\dot{\alpha}\alpha}$
where $\sigma_M=(i\vec{\tau}, 1)$ and
$\bar{\sigma}_M = (-i\vec{\tau},1)$, with Pauli matrices $\tau$.
All of these definitions follow the notation of \cite{Dorey:2002ik}. 
For the derivation of this matrix model via gauge/string duality, 
see \cite{Hashimoto:2010je}.

\subsection{Two-body nuclear forces.}

We review briefly \cite{Hashimoto:2010je} for explaining 
how to obtain the two-body nuclear forces. First, we describe a single
baryon wave function, and second, obtain the two-body Hamiltonian by
integrating out $A_0$ with a simple gauge choice. 

In all the cases, we
need to solve the ``ADHM constraint'' \cite{Atiyah:1978ri} 
which minimizes the 
potential induced by integrating out $D_s$. This is because the potential
has a coefficient $\lambda^2$ which is very large 
in the gauge/string duality.
\begin{eqnarray}
\vec{\tau}^{\;\dot{\alpha}}_{\;\;\;\dot{\beta}}
\left(
\bar{X}^{\dot{\beta}\alpha} X_{\alpha \dot{\alpha}}
+
\bar{w}^{\dot{\beta}}_{i} w_{\dot{\alpha}i} \right)_{BA}= 0\, .
\label{ADHM}
\end{eqnarray}
Here $A,B = 1, ..., k$. 

For a single baryon $k=1$, this equation is simply solved by
$w_{\dot{\alpha} i} = U_{\dot{\alpha}i}\rho$ where $U$ is an $SU(2)$
matrix and $\rho$ is a constant. After integrating out the auxiliary
field $A_0$, the matrix model action becomes a standard quantum
mechanics whose Lagrangian is almost the same as that of the soliton
approach \cite{Hong:2007kx, Hata:2007mb}, 
\begin{eqnarray}
S &=& 
\frac{\lambda N_c M_{\rm KK}}{54 \pi} 
\int\! dt \; {\rm tr}_k 
\left[
(\p_0 X^4)^2 -\frac23 M_{\rm KK}^2 (X^4)^2 
\right.
\nonumber \\
&& \left.
\hspace{20mm}
+ \p_0 \bar{w}^{\dot{\alpha}}_i \p_0 w_{\dot{\alpha}i} 
- \frac16  M_{\rm KK}^2 \bar{w}^{\dot{\alpha}}_i w_{\dot{\alpha}i}
-\left(\frac{27\pi}{\lambda M_{\rm KK}}\right)^2
\frac{1}{\bar{w}^{\dot{\alpha}}_i w_{\dot{\alpha}i}}
\right]\, .
\end{eqnarray}
This quantum mechanics is solved by following \cite{Hong:2007kx, Hata:2007mb}. 
At the leading order in the large $N_c$ limit, 
the wave functions for $X^4$ and $\rho$
are classical, which simply means that we can substitute the classical
values 
\begin{eqnarray}
 X^4 = 0\, , \quad 
\rho = 2^{-1/4}3^{7/4} \sqrt{\pi} \lambda^{-1/2} M_{\rm KK}^{-1}\, .
\label{classicalxr}
\end{eqnarray}
The wave functions for the spin/isospin $U$ is nontrivial. 
They are shared with those of the Skyrme model \cite{Skyrme:1962vh,ANW},
as described in \cite{Hata:2007mb}, and given by 
\begin{eqnarray}
\langle \vec{a}|\left(
\begin{array}{cc}
| p \uparrow \rangle & | p \downarrow \rangle
 \\
| n \uparrow \rangle & | n \downarrow \rangle
\end{array}
\right)_{IJ} \, 
= 
 \frac1{\pi} (\tau^2 U)_{IJ}
=
\frac1{\pi} 
\left(
\begin{array}{cc}
a_1 + i a_2 & -a_3-ia_4
 \\
-a_3 +i a_4 & -a_1 + i a_2 
\end{array}
\right)_{IJ} \, .
\label{wave}
\end{eqnarray}
The $SU(2)$ matrix $U$ is represented by a unit 4-vector $\vec{a}$ as
$U = i a_i \tau^i + a_4{\bf 1}_{2\times 2}$, with 
$(a_1)^2 + (a_2)^2 + (a_3)^2 + (a_4)^2 = 1$.

Next, let us review the case $k=2$ of \cite{Hashimoto:2010je}. 
The generic solution to the ADHM constraint (\ref{ADHM}) with 
$k=2, N_f=2$ is the well-known ADHM data of two $SU(2)$ YM instantons,
\begin{eqnarray}
 X^M = \tau^3 \frac{r^M}{2} + \tau^1 Y^M \, , 
\quad 
w^{A=1}_{\dot{\alpha}i}= U^{(A=1)}_{\dot\alpha i} \rho_1\, , 
\quad  w^{A=2}_{\dot{\alpha}i}= U^{(A=2)}_{\dot\alpha i} \rho_2\, .
\label{configx}
\end{eqnarray}
Here the off-diagonal part $Y$ is defined as 
\begin{eqnarray}
&& Y^M \equiv -\frac{\rho_1\rho_2}{4|r|^2}{\rm tr}
\left[
\bar{\sigma}_M r^N \sigma_N 
\left(
(U^{(1)})^\dagger U^{(2)}
-(U^{(2)})^\dagger U^{(1)}
\right)
\right] \, ,
\label{defY}
\end{eqnarray}
The vector $r^M$ ($M=1,2,3,4$)
is the distance between the two baryons, and $|r|^2 \equiv (r^M)^2$.
The $SU(2)$ matrices $U^{(1)}$ and $U^{(2)}$ 
denote the moduli parameters of each baryon, 
while $\rho_1$ and $\rho_2$ denote the 
moduli parameter associated with the size of instantons of each baryon. 
With this choice, the potential associated with $\vec{D}$ ({\it i.e.}
the ADHM potential) vanishes. 

We integrate out the auxiliary field $A_0$ to obtain the two-body
Hamiltonian. With the U(2) decomposition 
$A_0 = A_0^0 {\bf 1}_{2\times 2} + A_0^1 \tau^1 + A_0^2 \tau^2 
+ A_0^3 \tau^3$, it is straightforward to evaluate the terms including
the gauge field $A_0$ in the action,
\begin{eqnarray}
&& 
S_{\rm kin. + CS} \equiv
\frac{\lambda N_c M_{\rm KK}}{54 \pi} 
 \int\! dt \; 
{\rm tr}\left[
(D_0 X^M)^2
+ D_0 \bar{w}^{\dot{\alpha}}_i D_0 w_{\dot{\alpha}i} \right]
+  N_c \int \! dt \; {\rm tr}A_0
\nonumber \\
&& 
= 
\frac{\lambda N_c M_{\rm KK}}{54 \pi} 
\int\! dt \; 
\biggm[
2(A_0^1)^2 |r|^2 + 8 (A_0^3)^2 (Y^M)^2
+
2(\rho_1^2+\rho_2^2)\left(
(A_0^0)^2 + (A_0^1)^2 +(A_0^3)^2
\right)
\nonumber 
\\
&& \left.
\hspace{20mm}
+ 4 \rho_1 \rho_2 A_0^0 A_0^1 \; {\rm tr}
\left[U^{(1)\dagger} U^{(2)}\right]
+ 4 (\rho_1^2-\rho_2^2)A_0^0 A_0^3
+ \frac{108\pi}{\lambda M_{\rm KK}} A_0^0
\right]\, .
\end{eqnarray}
Solving the equations of motion for all the components of $A_0$
and substituting the solution back to this action, 
we obtain 
the potential $\int dt \; V = -S_{\rm kin.+CS}$, 
\begin{eqnarray}
&&V = 2 V_{\rm 1-body}+ V_{\rm 2-body}\, , 
\quad V_{\rm 1-body}
= \frac{27\pi N_c}{4\lambda M_{\rm KK}}
\frac{1}{\rho^2} \, ,
\\
&&
V_{\rm 2-body} 
= \frac{27\pi N_c}{\lambda M_{\rm KK}}
\frac{(u_0)^2}{|r|^2 + 2\rho^2 - 2 (u_0)^2 \rho^2}
\, .
\label{2bodya}
\end{eqnarray}
Here $u_0 \equiv (1/2){\rm tr}\left[U^{(1)\dagger}U^{(2)}\right]$, and
we put $\rho_1=\rho_2 (=\rho)$ which is justified as we keep only the
leading term in the large $N_c$ expansion. The value of 
$\rho$ is (\ref{classicalxr}).\footnote{
For our later purpose, we write the expression for the case 
of classically aligned spins and isospins.
This corresponds to $U^{(1)}=U^{(2)}$, which is nothing but an ADHM data
for 'tHooft instantons. The two-body potential is found as 
\begin{eqnarray}
V_{\rm 2-body}^{\rm cl} 
= \frac{27\pi N_c}{\lambda M_{\rm KK}}
\frac{1}{|r|^2}
\, .
\label{twobodyclassical}
\end{eqnarray}
}

In addition to the terms in $S_{\rm kin.+CS}$, there is the mass term 
for $X^4$ in the Lagrangian, 
\begin{eqnarray}
\frac{\lambda N_c M_{\rm KK}}{54 \pi}  \cdot \frac{2}{3} M_{\rm KK}^2
{\rm tr}(X^4)^2 =
\frac{\lambda N_c}{81 \pi} M_{\rm KK}^3
\left(
(r^4)^2/2 + 2(Y^4)^2
\right)\, .
\end{eqnarray}
The off-diagonal component $Y$ gives an additional two-body potential, 
\begin{eqnarray}
 V^{\rm mass}_{\rm 2-body} =   \frac{\lambda N_c M_{\rm KK}^3 }{162 \pi}
\frac{\rho^4}{|r|^4} \left(
r^i \; {\rm tr}
\left[i\tau^i \left(U^{(1)}\right)^\dagger U^{(2)}\right]
\right)^2
\, ,
\label{massv2}
\end{eqnarray}
where $i=1,2,3$.
So the 2-body potential is a sum of (\ref{2bodya}) and (\ref{massv2}).
The four-dimensional distance $|r|$ is equal to the inter-baryon
distance $|r^i|^2$ in three 
dimensions, since the classical value of the $X^4$ for the single
instantons is zero at the large $N_c$ leading order, 
as in (\ref{classicalxr}).

Using the nucleon wave function (\ref{wave}), it is straightforward to
evaluate the vacuum expectation value of this potential. The final form
of the two-body nuclear potential is
$ \langle V\rangle_{I_1,J_1,I_2,J_2}
= V_{\rm C}(\vec{r}) + S_{12} V_{\rm T}(\vec{r})$
with the standard definition
$S_{12} \equiv 12 J_1^i \hat{r}^i J_2^j \hat{r}^j - 4 J_1^i J_2^i$
(with $\hat{r}^i \equiv r^i/|r|, i=1,2,3$),
where the central and tensor forces are\footnote{
The result is quite close to that of the soliton approach
\cite{Hashimoto:2009ys}.}
\begin{eqnarray}
 V_{\rm C}(\vec{r})
=
\pi \left(
\frac{3^3}{2} + 8 I_1^i I_2^i J_1^j J_2^j
\right)
\frac{N_c}{\lambda M_{\rm KK}}
\frac{1}{|r|^2} \, ,
\quad
 V_{\rm T}(\vec{r})
= 2\pi I_1^i I_2^i
\frac{N_c}{\lambda M_{\rm KK}}
\frac{1}{|r|^2} \, .
\label{centen}
\end{eqnarray}


\vspace{2mm}
\noindent

\section{Three-body baryon interaction.}
\label{sec3}

The three-body interaction potential can be computed by using the matrix
model with $k=3$ for $k\times k$ matrices. 
The procedures to compute the nuclear potential
are parallel to the case of the two baryons in the previous
section, and here is a summary of the procedures:  
\begin{itemize}
 \item[(1)] Choose your $k$, and 
solve the ADHM constraint (\ref{ADHM}) (which minimizes the
	    potential obtained by integrating out the auxiliary field
	    $D$). 
\item[(2)]
Substitute the solution back to the matrix model Lagrangian. 
\item[(3)] Integrate out the 
auxiliary field $A_0$.
\item[(4)] Evaluate the Hamiltonian with your favorite baryon state. 
The baryon state is just a tensor product of single-baryon states
(which is given by the $k=1$ analysis).
\end{itemize}
In this section,
at first as an exercise, we consider a case where all
the three 
baryons share the same {\it classical} spin/isospins. Then for next, after giving an
explicit set-up for generic {\it quantum} spin/isospins, we demonstrate exact
computations for baryons aligned on a straight line with equal
spacings. The reason for choosing this linear position is just to
simplify and illuminate the computations.  
Finally we evaluate the
three-body Hamiltonian with 
specific three-baryon quantum states: (a)
three neutrons with spins averaged, and (b) proton-proton-neutron
and proton-neutron-neutron. 
We find that the three-body nuclear potential is positive {\it i.e.} repulsive.

\subsection{An exercise : classical treatment of spin/isospin.}
\label{sec3-1}

Let us evaluate the three-body Hamiltonian, first, for a simple
situation where all the baryons share the same classical spin/isospin,
as an exercise. 

\vspace{3mm}
\noindent
\underline{Procedure (1): solving the ADHM constraint}
\vspace{3mm}

First, let us consider the configuration space of minimizing the
$\vec{D}$ term. This is equivalent to the so-called ADHM constraints
(\ref{ADHM})
for any $A,B=1,2,3$. A simple solution to this constraint equation is
the ADHM data for 'tHooft instantons, which we treat in this subsection,
while in the later subsection we consider generic solution to this
constraint. The ADHM data for the 'tHooft instantons consists of
{\it diagonal} matrices $X$ and special $w$'s sharing the same
orientation, 
\begin{eqnarray}
w^A_{\dot \alpha i} 
=  U_{\dot \alpha i} \rho^A \quad (A = 1, 2, 3) \,,
\label{thooftw}
\end{eqnarray}
where $2\times 2$ unitary matrix $U$ is independent of the 
index $A$. As the degrees of freedom $w$ correspond to the spin and the
isospin, this means that all the three baryons share the same
``classical'' spins and isospins. Here, fixing the orientation $U$ for 
baryons cannot be achieved with wave functions with finite width, that
is the reason we call this ``classical.'' 
As the off-diagonal elements of $X^M$ vanish, all the commutators
$[X,X]$ are zero, which trivially satisfies (\ref{ADHM}). 

In \cite{Hashimoto:2009as}, 'tHooft instantons are used in the soliton
approach to evaluate the three-body nuclear forces. The result turns out
to vanish. In this subsection, we will find that our matrix model also
gives the same answer, the vanishing three-body force for the ADHM data
of the 'tHooft instantons.

We are going to choose implicitly the gauge
$\partial_0 w^A_{\dot{\alpha} i}  = 0 $ so that there is no
time-dependence in $w$. See \cite{Hashimoto:2010je} for details of
the gauge choices. 
The matrices $X^M$ whose diagonal elements with $M=1,2,3$
specify the spatial location of the 
baryons are diagonal,  
\begin{eqnarray}
 X^M = \sum_{a=3,8} {\lambda^a \over 2} r^M_a \,.
\label{threeconfigx}
\end{eqnarray}

\vspace{3mm}
\noindent
\underline{Procedure (2): substitute the ADHM data to the action}
\vspace{3mm}

The inter-baryon potential consists of two terms, the potential
coming from the integration of $A_0$, and the potential from the mass
term for $X^4$. The latter vanishes for the 'tHooft instantons, as there
is no off-diagonal extra component in (\ref{threeconfigx}). So in this
subsection we concentrate on the former.

Given the ADHM data, we can integrate out $A_0$,
in analogy to the two-body case. The auxiliary field 
$A_0$ is expanded by the Gell-Mann matrices $\lambda^a$,
\begin{eqnarray}
 A_0 = A_0^0 {\bf 1}_{3\times 3} + \sum_{a=1}^8
A_0^a {\lambda^a \over 2}\,. 
\end{eqnarray}
As in the two-body case, in the Lagrangian the terms containing $A_0$ are 
$(D_0X)^2$ and $D_0 \bar w D w$. 
The CS term contains only the overall $U(1)$ component, $A_0^0$.  

First, the kinetic terms of $X$ gives 
\begin{eqnarray}
 {\rm tr}(D_0 X^M)^2
&=&  \frac{1}{2}\left((A_0^1 r_3^M)^2 + (A_0^2 r_3^M)^2 \right) 
 +  
\frac{1}{8}\left((A_0^4)^2 + (A_0^5)^2 \right)(r_3^M + \sqrt{3} r_8^M)^2 
\nonumber 
\\
 & & 
+\frac{1}{8}\left((A_0^6)^2 + (A_0^7)^2 \right) (r_3^M - \sqrt{3}
r_8^M)^2  \,. 
\label{Xexpansion}
\end{eqnarray}
Next we consider the kinetic term for $w$,
\begin{eqnarray}
 {\rm tr}D_0 \bar{w}^{\dot{\alpha}}_i D_0 w_{\dot{\alpha}i} 
= 
\left(A_0^0\right)^2  
\left(\sum_A |w^A_{\dot\alpha i}|^2
\right)
 + 2 A_0^a A_0^0 \tilde{j}^a 
+ \tilde{t}^{ab} A_0^a A_0^b
\end{eqnarray}
where $a(=1,\cdots, 8)$ 
is the adjoint index of the $SU(3)$,
and $A (= 1,2,3)$ is the index for the baryons. 
The coefficients $\tilde{j}^a$ and $\tilde{t}^{ab}$ are defined as
\begin{eqnarray}
\tilde{j}^a \equiv w^A_{\dot \alpha i}  
\frac{\lambda^a_{AB}}{2} (w^B_{\dot \alpha i})^* \,, \quad 
\tilde{t}^{ab} \equiv w^A_{\dot \alpha i}  
{\lambda^a_{AB} \lambda^b_{BC} \over 4} (w^C_{\dot \alpha i})^* \,.
\end{eqnarray}
Using the 
definition of symmetric structure constants $d^{abc}$ for $SU(3)$, 
\begin{eqnarray}
\left\{\frac{\lambda^a}{2} \,, \frac{\lambda^b}{2}
\right\} 
= d^{abc} \frac{\lambda^c}{2} + \frac{1}{3} \delta^{ab} \,,
\end{eqnarray}
the term with $\tilde t^{ab}$ is replaced by 
\begin{eqnarray}
\label{tAA}
\tilde{t}^{ab}A_0^a A_0^b =  \frac{1}{6} w^B_{\dot \alpha i}  
 (w^B_{\dot \alpha i})^*  \delta^{ab} A_0^a A_0^b + 
\frac{d^{abc}}{4} w^A_{\dot \alpha i}  
\lambda^c_{AC} (w^C_{\dot \alpha i})^*   A_0^a A_0^b \,.
\end{eqnarray}
Now due to the ADHM data (\ref{thooftw}), all $w^A_{\dot
\alpha i}$ are proportional to each other, {\it i.e.}, 
$w^{A=1}_{\dot \alpha i}/\rho^{A=1} 
= w^{A=2}_{\dot \alpha i}/\rho^{A=2}
= w^{A=3}_{\dot \alpha i}/\rho^{A=3}$. 
Therefore, 
$w^A_{\dot \alpha i}  \lambda^c_{AC} (w^C_{\dot \alpha i})^*$ 
vanishes for $c=2,5,7$, and 
\begin{eqnarray}
\label{thsimpone}
\frac{w^A_{\dot \alpha i}  \lambda^{c=1}_{AC} 
(w^C_{\dot \alpha i})^*}{\rho^1 \rho^2} 
=  \frac{w^A_{\dot \alpha i}  \lambda^{c=4}_{AC} 
(w^C_{\dot \alpha i})^*}{\rho^1 \rho^3}
= \frac{w^A_{\dot \alpha i}  \lambda^{c=6}_{AC} 
(w^C_{\dot \alpha i})^*}{\rho^2 \rho^3} 
= 2 U_{\dot \alpha i} (U_{\dot \alpha i})^{\dagger} 
= 4\,. 
\end{eqnarray}
Furthermore, similar terms with $c=3$ and $c=8$ are given by 
\begin{eqnarray}
\label{thsimptwo}
{w^A_{\dot \alpha i}  \lambda^{c=3}_{AC} 
(w^C_{\dot \alpha i})^*}
= 
2 \left( (\rho^1)^2 -(\rho^2)^2 \right)   \,,\quad 
{w^A_{\dot \alpha i}  \lambda^{c=8}_{AC} (w^C_{\dot \alpha i})^*}
= 
\frac{2}{\sqrt{3}}\left( (\rho^1)^2 +(\rho^2)^2 - 2 (\rho^3)^2\right)\,.
\nonumber 
\end{eqnarray}
Using the symmetric structure constant $d^{abc}$, we get
\begin{eqnarray}
\tilde{t}^{ab}A_0^a A_0^b  &=& 
\frac{1}{6} w^B_{\dot \alpha i}   (w^B_{\dot \alpha i})^*  \delta^{ab}
A_0^a A_0^b 
+ \frac{d^{abc}}{4} w^A_{\dot \alpha i}  \lambda^c_{AC} (w^C_{\dot
\alpha i})^*   A_0^a A_0^b \nonumber \\ 
&=& \frac{1}{3} \left( \sum_{a=1}^8 (A_0^a)^2 \right) \left((\rho^1)^2 +
		 (\rho^2)^2 + (\rho^3)^2 \right)  
+ \left( \frac{2}{\sqrt{3}} A_0^1 A^8_0 + A_0^4 A_0^6 + A_0^5 A_0^7
  \right) \rho^1 \rho^2  
\nonumber \\&& 
+ \left( - \frac{1}{\sqrt{3}} A_0^4 A^8_0 + A_0^1 A_0^6 -  A_0^2 A_0^7 +
   A_0^3 A_0^4 \right) \rho^1 \rho^3  
\nonumber \\
&&
+ \left( - \frac{1}{\sqrt{3}} A_0^6 A^8_0 + A_0^1 A_0^4 + A_0^2 A_0^5 -
   A_0^3 A_0^6 \right) \rho^2 \rho^3 \nonumber \\ 
&& + \left( \frac{1}{\sqrt{3}} A_0^3 A_0^8 +  \frac{1}{4} (A_0^4)^2 +
      \frac{1}{4} (A_0^5)^2 -  \frac{1}{4} (A_0^6)^2 -  \frac{1}{4}
      (A_0^7)^2  \right) \left( (\rho^1)^2 - (\rho^2)^2 \right)
      \nonumber \\ 
&& + \left(   \frac{1}{6} (A_0^1)^2 +  \frac{1}{6} (A_0^2)^2 +
      \frac{1}{6} (A_0^3)^2 -  \frac{1}{6} (A_0^8)^2  \right) \left(
      (\rho^1)^2 + (\rho^2)^2  - 2 (\rho^3)^2\right) \nonumber \\ 
&& + \left(   -\frac{1}{12} (A_0^4)^2 -  \frac{1}{12} (A_0^5)^2 -
      \frac{1}{12} (A_0^6)^2 -  \frac{1}{12} (A_0^7)^2  \right) \left(
      (\rho^1)^2 + (\rho^2)^2  - 2 (\rho^3)^2\right) \,.
\nonumber \\
&&  
\end{eqnarray} 
Next, we consider  
$2 A_0^a A_0^0 \tilde{j}^a $. Again, due to the simplicity of the
ADHM data for the 't Hooft instantons, 
$\tilde j^a$ is nonzero only for $a = 1,4,6, 3,8$, and we obtain
\begin{eqnarray}
2 A_0^a A_0^0 \tilde{j}^a &=& 4 A_0^1 A_0^0 \rho^1 \rho^2 + 4 A_0^4
 A_0^0 \rho^1 \rho^3  
+4 A_0^6 A_0^0 \rho^2 \rho^3 
\nonumber \\
&& + 2 A_0^3 A_0^0 \left( (\rho^1)^2 - (\rho^2)^2 \right) +
\frac{2}{\sqrt{3}}A_0^8 A_0^0 \left( (\rho^1)^2 + (\rho^2)^2  - 2
			       (\rho^3)^2\right) \,.
\end{eqnarray}
In addition, we have 
$\left(A_0^0\right)^2  
\left(\sum_A |w^A_{\dot\alpha i}|^2
\right) = 2 \left((\rho^1)^2 + (\rho^2)^2 + (\rho^3)^2 \right)
\left(A_0^0\right)^2$.
So, in total, the kinetic term for $w$ is evaluated as 
\begin{eqnarray}
\lefteqn{ {\rm tr}D_0 \bar{w}^{\dot{\alpha}}_i D_0 w_{\dot{\alpha}i} }
\nonumber \\
&&
=2 \left((\rho^1)^2 + (\rho^2)^2 + (\rho^3)^2 \right)\left(
    \left(A_0^0\right)^2  
+ \frac{1}{6} \sum_{a=1}^8 (A_0^a)^2 \right)  
+
4\rho^1 \rho^2  A_0^1 A_0^0 + 4  \rho^1 \rho^3 A_0^4 A_0^0 
\nonumber \\&& 
+ 4 \rho^2 \rho^3  A_0^6 A_0^0 
+ 2 A_0^3 A_0^0 \left( (\rho^1)^2 - (\rho^2)^2 \right) +
\frac{2}{\sqrt{3}}A_0^8 A_0^0 \left( (\rho^1)^2 + (\rho^2)^2  - 2
			       (\rho^3)^2\right)  
\nonumber \\ 
&& 
+  \frac{2 \rho^1 \rho^2}{ \sqrt{3}} A_0^1 A^8_0 +{\rho^1 \rho^2} A_0^4
A_0^6 + {\rho^1 \rho^2} A_0^5 A_0^7   
- \frac{ \rho^1 \rho^3}{\sqrt{3}} A_0^4 A^8_0 +{ \rho^1 \rho^3} A_0^1
A_0^6 - { \rho^1 \rho^3} A_0^2 A_0^7 
\nonumber \\ 
&& 
+ { \rho^1 \rho^3} A_0^3 A_0^4    
- \frac{\rho^2 \rho^3}{ \sqrt{3}} A_0^6 A^8_0 + {\rho^2 \rho^3} A_0^1
A_0^4 + {\rho^2 \rho^3} A_0^2 A_0^5 - {\rho^2 \rho^3} A_0^3 A_0^6
\nonumber  \\ && 
+ \left( \frac{1}{\sqrt{3}} A_0^3 A_0^8 +  \frac{1}{4} (A_0^4)^2 +
   \frac{1}{4} (A_0^5)^2 -  \frac{1}{4} (A_0^6)^2 -  \frac{1}{4}
   (A_0^7)^2  \right) \left( (\rho^1)^2 - (\rho^2)^2 \right) 
\nonumber \\ && 
+ \frac{1}{12}\left(2 (A_0^1)^2 +2 (A_0^2)^2 + 2
   (A_0^3)^2 - 2 (A_0^8)^2   
 - (A_0^4)^2 -  (A_0^5)^2 - 
 (A_0^6)^2 -  (A_0^7)^2   
\right) 
\nonumber \\ && 
\times \left( (\rho^1)^2 + (\rho^2)^2  - 2 (\rho^3)^2\right) \,. 
\label{wexpansion}
\end{eqnarray}
Finally, the CS term has only the  $A_0^0$ element,
\begin{eqnarray}
L_{\rm CS} = \frac{162 \pi}{\lambda M_{KK}} A_0^0 \,.
\label{CSAzero}
\end{eqnarray}
The total action $L_{A_0}$ involving the gauge field $A_0$
is a sum of  (\ref{Xexpansion}),
(\ref{wexpansion}) and (\ref{CSAzero}), as 
\begin{eqnarray}
L_{A_0} \equiv {\rm tr}(D_0 X^M)^2   + {\rm tr}D_0
 \bar{w}^{\dot{\alpha}}_i D_0 w_{\dot{\alpha}i}  + L_{\rm CS}
\label{LAzero}
\end{eqnarray}

\vspace{3mm}
\noindent
\underline{Procedure (3): integrate out $A_0$}
\vspace{3mm}

We have to solve the simultaneous equations for all $A_0^a$ and $A_0^0$, 
\begin{eqnarray}
\frac{\partial L_{A_0}}{\partial A_0^0} = \frac{\partial
 L_{A_0}}{\partial A_0^a} = 0  \quad 
\mbox{(for all $a = 1, \cdots ,  8$)} 
\label{simleqs}
\end{eqnarray}
Although $A_0^0$ is mixed  with the other components $A_0^a$
a unique solution is found as 
\begin{eqnarray}
&&A^1_0 =  \frac{27 \pi}{\lambda M_{KK}} \frac{(\rho^1)^2 +
(\rho^2)^2}{(r_3^M)^2 \rho^1 \rho^2} \,, \quad
A^4_0 =  \frac{108 \pi}{\lambda M_{KK}} \frac{(\rho^1)^2 +
(\rho^3)^2}{(r_3^M + \sqrt{3} r_8^M )^2 \rho^1 \rho^3} \,, 
\nonumber\\&&
A^6_0 =  \frac{108 \pi}{\lambda M_{KK}} 
\frac{(\rho^2)^2 + (\rho^3)^2}{(r_3^M - \sqrt{3} r_8^M )^2 \rho^2
\rho^3} \,, \quad
A^2_0 = A^5_0 = A^7_0 = 0 \,,
\end{eqnarray}
$A_0^0$, $A_0^3$ and $A_0^8$ have complicated expressions, so we
omit to write them explicitly here. 
We plug the solution back to the action 
$L_{A_0}$ given by (\ref{LAzero}), then we obtain integrated action
$L_{A_0}$ in terms of the moduli parameters 
$r_3^M$,  $r_8^M$ and $\rho$, as  
\begin{eqnarray}
\lefteqn{L_{A_0} \left(r_3^M, r_8^M, \rho^A \right) }
\nonumber \\
&&
= \left( \frac{54 \pi}{\lambda M_{KK}}  \right)^2 
\left[-\sum_{A=1}^3 \frac{1}{8 (\rho^A)^2}  -\frac{1}{4 (r_3^M)^2}
 \left( 1 + \frac{(\rho^1)^2}{2 (\rho^2)^2} + \frac{(\rho^2)^2}{2
  (\rho^1)^2}\right) 
\right. \nonumber \\
&& 
\hspace{30mm}
-\frac{1}{ (r_3^M + \sqrt{3} r_8^M)^2} 
\left( 1 + \frac{(\rho^1)^2}{2 (\rho^3)^2} +
 \frac{(\rho^3)^2}{2 (\rho^1)^2}\right) 
\nonumber \\
&& \hspace{30mm}
\left. 
-\frac{1}{ (r_3^M - \sqrt{3} r_8^M)^2} 
\left( 1 + \frac{(\rho^2)^2}{2 (\rho^3)^2}
 + \frac{(\rho^3)^2}{2 (\rho^2)^2}\right)
\right] \,.
\end{eqnarray}
The total Hamiltonian (potential) $V^{\rm cl}$ 
is given by 
\begin{eqnarray}
S = \frac{\lambda N_c M_{KK}}{54 \pi} \int\! dt \; L_{A_0} 
\equiv - \int\! dt \;V^{\rm cl} \,,
\end{eqnarray}
as in the two-body case. We obtain
\begin{eqnarray}
\label{thooftVpot}
V^{\rm cl} 
&=& \left( \frac{54 \pi N_c}{\lambda M_{KK}}
			       \right)  
\left[\sum_{A=1}^3 \frac{1}{8 (\rho^A)^2}  
+\frac{1}{4 (r_3^M)^2} \left( 1 + \frac{(\rho^1)^2}{2 (\rho^2)^2} 
+ \frac{(\rho^2)^2}{2 (\rho^1)^2}\right)
\right. \nonumber \\
&& 
\hspace{20mm}
+\frac{1}{(r_3^M + \sqrt{3} r_8^M)^2} 
\left( 1 + \frac{(\rho^1)^2}{2 (\rho^3)^2} 
+ \frac{(\rho^3)^2}{2 (\rho^1)^2}\right)
\nonumber \\
&& 
\left. 
\hspace{20mm}
+\frac{1}{ (r_3^M - \sqrt{3} r_8^M)^2} 
\left( 1 + \frac{(\rho^2)^2}{2 (\rho^3)^2} 
+ \frac{(\rho^3)^2}{2 (\rho^2)^2}\right)
\right] \,.
\end{eqnarray}

To find the potential intrinsic to the three-body, we need to subtract
the one-body and two-body Hamiltonians. For the ADHM data for the
'tHooft instantons, they are given by \cite{Hashimoto:2010je}
\begin{eqnarray}
 V_{\rm 1-body}^{\rm cl} 
= \frac{27\pi N_c}{4\lambda M_{\rm KK}}\frac{1}{\rho^2}
\, , \quad 
V_{\rm 2-body}^{\rm cl} = \frac{27\pi N_c}{4 \lambda M_{\rm KK}}
\left(2 + \frac{\rho_2^2}{\rho_1^2} + \frac{\rho_1^2}{\rho_2^2}\right)
\frac{1}{(r^M)^2}\, , 
\end{eqnarray}
where $r^M$ is the distance between the two baryons. The subtraction of
these give
\begin{eqnarray}
 V^{\rm cl} - \sum_{A=1,2,3} V_{\rm 1-body}^{\rm (A),cl}
-\frac12 \sum_{A\neq B} V_{\rm 2-body}^{\rm (A,B),cl}
= 0.
\end{eqnarray}
Therefore, 
the three-body forces vanish, for the baryons sharing the same classical
spin/isospins. This result is the same as 
the one given in the soliton approach \cite{Hashimoto:2009as}.   

The ``classical'' spins and isospins are realized when the magnitude of
the spin/isospins is large, which is only possible for heavy higher spin
baryons, but not for  
spin $1/2$ nucleons. Therefore unfortunately 
this ``classical'' treatment does not work for the realistic nucleons. 
Next, we keep the quantum spin/isospin degrees of
freedom (the phase in $w$) explicitly in the computation and provide 
a framework for nuclear forces with standard quantum spin/isospins.

\subsection{Generic three-body interactions: a set-up.}
\label{sec3-2}

\vspace{3mm}
\noindent
\underline{Procedure (1): solving the ADHM constraint}
\vspace{3mm}

The computations with the ADHM data for the 'tHooft instantons are easy
but they are not realistic system, since the spin/isospin rotation
matrix $U$ is fixed by 
hand. We have to allow arbitrary $U$ for each baryon, in general.  
This means, instead of the previous (\ref{thooftw}),
we allow\footnote{Note that in single instanton
case, we have gauge freedom from $A_0$ to chose this $\partial_0w = 0$
gauge. In three instanton case, we have $A_0^0$, $A_0^3$, and $A_0^8$
gauge freedom to choose this $\partial_0 w^{A=1} = \partial_0 w^{A=2} = \partial_0 w^{A=3} = 0$.}
\begin{eqnarray}
w^A_{\dot \alpha i} =  U^A_{\dot \alpha i} \rho^A  
\quad (A = 1, 2, 3) \,.
\end{eqnarray}
In order to satisfy the ADHM constraint (\ref{ADHM}) with this generic
$w$, the off-diagonal components of the matrices $X^M$ should be turned
on, instead of (\ref{threeconfigx}),  
\begin{eqnarray}
X^M = \sum_{a=3,8} {\lambda^a \over 2} r^M_a + 
\sum_{a=1,4,6} {\lambda^a \over 2} r^M_a\, .
\label{rdef}
\end{eqnarray}
The diagonal $r_3$ and $r_8$ specify the positions of the three
baryons, while the off-diagonal $r_1, r_4$ and $r_6$ are small.

Although
generic three-instanton ADHM data is not available, we may need only the
ADHM data for well-separated instantons, 
\begin{eqnarray}
|r_3+ \sqrt{3} r_8|/2\, , \,
|-r_3+ \sqrt{3} r_8 |/2\, , \,
|r_8| \gg \rho \, ,
\end{eqnarray}
since the classical size of the instanton (baryon) is quite small as 
$\rho \sim 1/\sqrt{\lambda}$ for large $\lambda$. The ADHM data for the well-separated
instantons is described in \cite{Christ:1978jy}. In our notation, it
is\footnote{
Our $r_1/2$, $r_4/2$ and $r_6/2$ correspond to $b_{12}$, $b_{13}$, and
$b_{23}$ of \cite{Christ:1978jy}, as they are the off-diagonal elements
of the matrix $X^M$. Our $\rho_i U^i$ corresponds to $q_i$ of
\cite{Christ:1978jy}. Our formulas (\ref{cd1}), (\ref{cd2}), and
(\ref{cd2}) can be obtained explicitly from Eq.~(5.13) of
\cite{Christ:1978jy}, by substituting recursively the expression of
$b_{ij}$. }  
\begin{eqnarray}
 r_1^M \sigma_M
&=& \frac{d_{12}^M \sigma_M}{|d_{12}|^2} \rho_1 \rho_2 
\left((U^2)^\dagger U^1 - (U^1)^\dagger U^2\right)
\nonumber 
\\
& & + 
\frac{\rho_1 \rho_2 \rho_3^2
d_{12}^M \sigma_M}{4|d_{12}|^2|d_{13}|^2 |d_{23}|^2} 
\left[
\left((U^3)^\dagger U^2 - (U^2)^\dagger U^3\right)d_{32}^\dagger d_{31}
\left((U^1)^\dagger U^3 - (U^3)^\dagger U^1\right)
\right.
\nonumber \\
& &  -
\left.
\left((U^3)^\dagger U^1 - (U^1)^\dagger U^3\right)d_{31}^\dagger d_{32}
\left((U^2)^\dagger U^3 - (U^3)^\dagger U^2\right)
\right]\,  + {\cal O}(1/d^5)\, ,
\label{cd1}
\end{eqnarray}
\begin{eqnarray}
 r_4^M \sigma_M
&=& \frac{d_{13}^M \sigma_M}{|d_{13}|^2} \rho_1 \rho_3 
\left((U^3)^\dagger U^1 - (U^1)^\dagger U^3\right)
\nonumber 
\\
& & +
\frac{\rho_1 \rho_3 \rho_2^2
d_{13}^M \sigma_M}{4|d_{12}|^2|d_{13}|^2 |d_{23}|^2} 
\left[
\left((U^2)^\dagger U^3 - (U^3)^\dagger U^2\right)d_{23}^\dagger d_{21}
\left((U^1)^\dagger U^2 - (U^2)^\dagger U^1\right)
\right.
\nonumber \\
& & -
\left.
\left((U^2)^\dagger U^1 - (U^1)^\dagger U^2\right)d_{21}^\dagger d_{23}
\left((U^3)^\dagger U^2 - (U^2)^\dagger U^3\right)
\right]\,  + {\cal O}(1/d^5)\, ,
\label{cd2}
\\
 r_6^M \sigma_M
&=& \frac{d_{23}^M \sigma_M}{|d_{23}|^2} \rho_2 \rho_3 
\left((U^3)^\dagger U^2 - (U^2)^\dagger U^3\right)
\nonumber 
\\
& & +
\frac{\rho_2 \rho_3 \rho_1^2
d_{23}^M \sigma_M}{4|d_{12}|^2|d_{13}|^2 |d_{23}|^2} 
\left[
\left((U^1)^\dagger U^3 - (U^3)^\dagger U^1\right)d_{13}^\dagger d_{12}
\left((U^2)^\dagger U^1 - (U^1)^\dagger U^2\right)
\right.
\nonumber \\
& & -
\left.
\left((U^1)^\dagger U^2 - (U^2)^\dagger U^1\right)d_{12}^\dagger d_{13}
\left((U^3)^\dagger U^1 - (U^1)^\dagger U^3\right)
\right]\,  + {\cal O}(1/d^5)\, .
\label{cd3}
\end{eqnarray}
Here we have defined 
\begin{eqnarray}
 d_{ij} \equiv d_{ij}^M\sigma_M 
\end{eqnarray}
where $d_{ij}$ is the distance vector between the $i$-th and the $j$-th
instantons. 
From (\ref{rdef}), the location of the first, second, 
and third instanton is 
\begin{eqnarray}
r^M = r_3^M/2 + r_8^M/2\sqrt{3} \,, 
-r_3^M/2 + r_8^M/2\sqrt{3} \,, -r_8^M/\sqrt{3} \,, 
\end{eqnarray} 
respectively. 
Therefore we have 
\begin{eqnarray}
&& d_{12} = -d_{21} = r_3\, , \, \\
&& d_{13}= -d_{31}= (r_3-  \sqrt{3} r_8)/2\, , \,\\
&& d_{23} = -d_{32}= (-r_3- \sqrt{3} r_8)/2\, .
\end{eqnarray}

\vspace{3mm}
\noindent
\underline{Procedure (2): substitute the ADHM data to the action}
\vspace{3mm}

As all $U^A$ matrices are different, we need to consider 
$w^A_{\dot \alpha i}  \lambda^c_{AC} (w^C_{\dot \alpha i})^* $ 
for all $c = 1, \cdots, 8$. 
However, because  $U^A_{\dot \alpha i}\in$ SU(2), 
$w^A_{\dot \alpha i}  \lambda^c_{AC} (w^C_{\dot \alpha i})^* $ for 
$c = 2, 5, 7$ vanish as they are proportional to 
$U^A_{\dot \alpha i} (U^B_{\dot \alpha i})^\dagger 
- U^B_{\dot \alpha i} (U^A_{\dot \alpha i})^\dagger$ 
with $A,B=1,2,3$. The other components are calculated as follows: 

\begin{eqnarray}
w^A_{\dot \alpha i}  \lambda^1_{AC} (w^C_{\dot \alpha i})^* &=& \rho^1
 \rho^2 \left( U^1_{\dot \alpha i} (U^2_{\dot \alpha i})^\dagger +
	 U^2_{\dot \alpha i} (U^1_{\dot \alpha i})^\dagger \right)  
=  4  \rho^1 \rho^2 \, u^{(12)}_0  \nonumber \\
w^A_{\dot \alpha i}  \lambda^3_{AC} (w^C_{\dot \alpha i})^* 
&=& 2 \left( (\rho^1)^2 - (\rho^2)^2 \right) \nonumber \\
w^A_{\dot \alpha i}  \lambda^4_{AC} (w^C_{\dot \alpha i})^* 
&=& \rho^1 \rho^3 \left( U^1_{\dot \alpha i} (U^3_{\dot \alpha
		   i})^\dagger + U^3_{\dot \alpha i} (U^1_{\dot \alpha
		   i})^\dagger \right) = 4  \rho^1 \rho^3 \,
u^{(13)}_0\nonumber \\ 
w^A_{\dot \alpha i}  \lambda^6_{AC} (w^C_{\dot \alpha i})^* 
&=& \rho^2 \rho^3 \left( U^2_{\dot \alpha i} (U^3_{\dot \alpha
		   i})^\dagger + U^3_{\dot \alpha i} (U^2_{\dot \alpha
		   i})^\dagger \right)= 4  \rho^2 \rho^3 \, u^{(23)}_0
\nonumber \\ 
w^A_{\dot \alpha i}  \lambda^8_{AC} (w^C_{\dot \alpha i})^* 
&=& \frac{2}{\sqrt{3}}\left( (\rho^1)^2 +(\rho^2)^2 - 2
		       (\rho^3)^2\right)  \,. 
\end{eqnarray}
Here we have used the fact that the $SU(2)$  
matrices $U^A_{\dot \alpha i} (U^B_{\dot \alpha i})^\dagger$ 
for $A \neq B, (A,B) = 1,2,3$ can be written as 
$u_0{\bf 1}_{2 \times 2}+ i \sum_{i=1}^3 u_i \tau^i$ with 
$\sum_{i=0}^3 (u_i)^2 = 1$,  
in terms of Pauli matrices $\tau^i$, {\it i.e.}
\begin{eqnarray}
U^A_{\dot \alpha i} (U^B_{\dot \beta i})^\dagger \equiv  u^{(AB)}_0
 ({\bf 1}_{2 \times 2})_{\dot \alpha \dot \beta}+ i \sum_{i=1}^3
 u^{(AB)}_i \tau^i_{\dot \alpha \dot \beta} \,. 
\end{eqnarray}
The definition of $u_0$ follows that of the two-baryon case,
(\ref{2bodya}).

In short, compared with the previous ADHM data for the 'tHooft
instantons, we have new parameters $r_a^M$ with $a=1,4,6$, and 
$u^{AB}_0$. 
We first describe the integration of $A_0$, followed by the
explanation of the potential due to the mass term ${\rm tr}[(X^4)^2]$.

Let us write the terms including $A_0$ in the matrix model action
explicitly. They are the kinetic terms of $X$ and $w$, and the CS term. 
Note that due to the fact that 
$w^A_{\dot \alpha i}  \lambda^c_{AC} (w^C_{\dot \alpha i})^*  = 0 $
again for $c = 2,5,7$, the calculation for the $w$ kinetic term 
$Dw (Dw)^*$ is very similar to that of the ADHM data for the 'tHooft 
instantons.  On the other hand, the kinetic term for
$X$ contains terms in (\ref{Xexpansion}) as well as 
terms including $r_a$ with $a=1,4,6$. 
\begin{eqnarray}
{\rm tr}(D_0 X^M)^2 = {\rm tr} 
\left( -i \left[A_0, \sum_{\zeta=1,3,4,6,8}
\frac{\lambda^a}{2} r^M_a\right] \right)^2  \, .
\end{eqnarray}
Due to the fact that terms including 
$A_0^a$ with $a = 2,5,7$ decouple from the 
terms including $A_0^b$ with $b = 1,3,4,6,8$,  
and the fact that all $A_0^a$ with $a = 2,5,7$ 
appear in the Lagrangian as quadratic terms, 
the equations of motion for $A_0^a$ with $a = 2,5,7$ are simply solved
by $A_0^a = 0$ $(a = 2,5,7)$. 
With this observation, the kinetic term for $X^M$ is simplified as 
\begin{eqnarray}
\lefteqn{{\rm tr}(D_0 X^M)^2}
\nonumber \\
 &&= 
\frac{1}{8} \left( (A_0^4)^2 r_1^2 + (A_0^6)^2 r_1^2 
+ 4 (A_0^1)^2 r_3^2 + (A_0^4)^2 r_3^2 + (A_0^6)^2 r_3^2 
- 2 A_0^1 A_0^4 r_1 r_4 
\right. \nonumber \\
 &&\left.  
- 2 \sqrt{3} A_0^6 A_0^8 r_1 r_4 + 6 A_0^1 A_0^6 r_3 r_4 
- 
   2 \sqrt{3} A_0^4 A_0^8 r_3 r_4 
+ (A_0^1)^2 (r_4)^2 + (A_0^6)^2 (r_4)^2 
\right. \nonumber \\
 &&\left.  
+ 3 (A_0^8)^2 (r_4)^2 
-   2 A_0^1 A_0^6 r_1 r_6 - 2 \sqrt{3} A_0^4 A_0^8 r_1 r_6 
- 6 A_0^1 A_0^4 r_3 r_6 + 
   2 \sqrt{3} A_0^6 A_0^8 r_3 r_6 
\right. \nonumber \\
 &&\left.  
- 2 A_0^4 A_0^6 r_4 r_6 
+ 4 \sqrt{3} A_0^1 A_0^8 r_4 r_6 \right. \nonumber \\
   && \left. + 
   (A_0^1)^2 (r_6)^2 + (A_0^4)^2 (r_6)^2 + 3 (A_0^8)^2 (r_6)^2 + 
   (A_0^3)^2 (4 (r_1)^2 + (r_4)^2 + (r_6)^2) \right. \nonumber \\
  && \left. + 4 \sqrt{3} A_0^4 A_0^6 r_1 r_8 + 
   2 \sqrt{3} (A_0^4)^2 r_3 r_8 - 2 \sqrt{3} (A_0^6)^2 r_3 r_8 - 
   2 \sqrt{3} A_0^1 A_0^6 r_4 r_8
       \right. \nonumber \\
   && \left.  - 6 A_0^4 A_0^8 r_4 r_8 - 2 \sqrt{3} A_0^1 A_0^4 r_6 r_8 - 
   6 A_0^6 A_0^8 r_6 r_8 
   + 3 (A_0^4)^2 r_8^2 + 3 (A_0^6)^2 r_8^2  
    \right. \nonumber \\
   && \left.
    - 2 A_0^3 (4 A_0^1 r_1 r_3 + 3 A_0^6 r_1 r_4 + A_0^4 r_3 r_4 - \sqrt{3} A_0^8 (r_4)^2 \right.\nonumber \\&& \left. - 
      3 A_0^4 r_1 r_6 + A_0^6 r_3 r_6 + \sqrt{3} A_0^8 (r_6)^2 + \sqrt{3} A_0^4 r_4 r_8 - 
      \sqrt{3} A_0^6 r_6 r_8) \right) \, .
\end{eqnarray}


The kinetic term for $w$ is similar to the previous case with the
'tHooft instanton ADHM data.
\begin{eqnarray}
\lefteqn{ {\rm tr}D_0 \bar{w}^{\dot{\alpha}}_i D_0 w_{\dot{\alpha}i} }
\nonumber \\
&&= 
2 \left((\rho^1)^2 + (\rho^2)^2 + (\rho^3)^2 \right)\left(
   \left(A_0^0\right)^2  
+ \frac{1}{6} \sum_{\eta=1,3,4,6,8} (A_0^\eta)^2 \right) 
+4\rho^1 \rho^2 u^{(12)}_0 A_0^1 A_0^0 
\nonumber \\
&& 
+ 4  \rho^1 \rho^3 u^{(13)}_0 A_0^4 A_0^0 +4 \rho^2 \rho^3  u^{(23)}_0
A_0^6 A_0^0 
+ 2 A_0^3 A_0^0 \left( (\rho^1)^2 - (\rho^2)^2 \right) 
\nonumber \\ && 
+  \frac{2}{\sqrt{3}}A_0^8 A_0^0 \left( (\rho^1)^2 + (\rho^2)^2  - 2
				  (\rho^3)^2\right) 
+  \frac{2 \rho^1 \rho^2 u^{(12)}_0}{ \sqrt{3}} A_0^1 A^8_0 + \rho^1
\rho^2 u^{(12)}_0 A_0^4 A_0^6  
\nonumber \\
&& 
- \frac{ \rho^1 \rho^3 u^{(13)}_0}{\sqrt{3}} A_0^4 A^8_0 + \rho^1 \rho^3
u^{(13)}_0 A_0^1 A_0^6  + \rho^1 \rho^3 u^{(13)}_0  A_0^3 A_0^4
\nonumber \\ 
&& - \frac{\rho^2 \rho^3 u^{(23)}_0 }{\sqrt{3}} A_0^6 A^8_0 + \rho^2
\rho^3 u^{(23)}_0 A_0^1 A_0^4 - \rho^2 \rho^3 u^{(23)}_0 A_0^3 A_0^6
\nonumber \\ 
&& + \left( \frac{1}{\sqrt{3}} A_0^3 A_0^8 +  \frac{1}{4} (A_0^4)^2  -
      \frac{1}{4} (A_0^6)^2  \right) \left( (\rho^1)^2 - (\rho^2)^2
				     \right) \nonumber \\ 
&& + \frac{1}{12}\left(2 (A_0^1)^2  +  2 (A_0^3)^2 -
     2 (A_0^8)^2    - (A_0^4)^2  -  
      (A_0^6)^2\right) \left( (\rho^1)^2 + (\rho^2)^2  - 2
      (\rho^3)^2\right)  \, . \quad
\end{eqnarray}
With the CS term given by (\ref{CSAzero})
the total Lagrangian is again written in the form (\ref{LAzero}).
Again, we have to solve the simultaneous equations (\ref{simleqs}).

Next, we shall describe the potential due to the mass term of the matrix
model, 
\begin{eqnarray}
 \frac{\lambda N_c M^3_{\rm KK}}{54\pi}\frac23 {\rm tr}
  \left[(X^4)^2\right]
=
 \frac{\lambda N_c M^3_{\rm KK}}{3^4\pi} 
\left[
\frac14(r_3^4+ r_8^4/\sqrt{3})^2
+\frac14(-r_3^4+ r_8^4/\sqrt{3})^2
+ \frac13 (r_8^4)^2
\right.
\nonumber \\
\hspace{20mm}
\left.
+\frac12 
\left(
(r_1^4)^2 + (r_2^4)^2 + (r_4^4)^2 + (r_5^4)^2 + (r_6^4)^2 + (r_7^4)^2
\right)
\right] \, . \qquad 
\label{massx4exp}
\end{eqnarray}
The first three terms correspond to the square of the diagonal elements $X^{4}_{ii}$ for $i= 1,2,3$, so  
these correspond to the three copies of the one-baryon
potential. The terms in the second line are the two-body and the
three-body terms. To evaluate these, we need explicit expressions for
the off-diagonal $r_1$, $r_2$, $r_4$, $r_5$, $r_6$ and $r_7$.

In principle, it is a straightforward calculation to determine the
three-body force from this. However the actual calculation turns out to
be extremely messy, and it is hard to get a physical interpretation from  
that. To extract the physical essence, next we will choose a particular 
alignment of the baryons to simplify the expression, so that final
answer is easier to analyze.  

\subsection{Three-body Hamiltonian for baryons aligned on a line.}
\label{sec3-3}
We will find that the Hamiltonian is simplified significantly when all
the baryons are aligned on a line. 
We consider the following case
\begin{eqnarray}
\label{rthree}
r^M_8 = 0 \,, \quad r_3^M \equiv r^M \neq 0 \, .
\end{eqnarray} 
This means that the first, the second and the third baryon are placed at 
$x^M = r^{M}_3/2$, $x^M = - r^{M}_3/2$,  and  $x = 0$, respectively.  
In this case, since $r^M_8 = 0$, the expression for the potential is 
significantly simplified. Still, since we are treating the  
spin/isospin moduli quantum mechanically, 
this gives lots of information on the three-body forces.  
In addition, 
we notice that all the size moduli $\rho_i$ can be taken to be a
classical value, 
$\rho_1 = \rho_ 2 = \rho_3 = \rho$, 
since we are dealing with only the leading order in the large $N_c$
limit. The large $N_c$ limit is the same as classical limit since the action (\ref{mm}) has 
overall $N_c$. 
This simplifies the computation too.

The resultant Lagrangian
concerning the gauge field $A_0$  is
\begin{eqnarray}
L_{A_0}= \frac{\lambda M_{\rm KK} N_c}{54 \pi}
\left(L_1 + L_2\right)
\end{eqnarray}
where 
\begin{eqnarray}
L_1& \equiv &
\frac{162 A_0^0 \pi }{\lambda  M_{\rm KK}}
+\frac{(A_0^1)^2  r^2}{2}
+\frac{(A_0^4)^2 r^2}{8}
+\frac{(A_0^6)^2 r^2}{8}
\nonumber \\
&& +
 \left(6 (A_0^0)^2+(A_0^1)^2+(A_0^3)^2
+(A_0^4)^2+(A_0^6)^2+(A_0^8)^2\right)\rho^2
\nonumber
\\
&& 
+ \left(A_0^4A_0^6
+\frac{2A_0^1 A_0^8}{\sqrt{3}}\right) \rho^2  u_0^{(12)}
+
\left(A_0^3 A_0^4
+A_0^1 A_0^6
-\frac{A_0^4A_0^8}{ \sqrt{3}}\right) \rho^2  u_0^{(13)}
\nonumber
\\
&& 
+\left(A_0^1 A_0^4-\!A_0^3A_0^6
-\frac{A_0^6A_0^8}{ \sqrt{3}}\right)\rho^2  u_0^{(23)}
 \!\!+\!4 A_0^0 \left(A_0^1   u_0^{(12)}
\!\!+ \! A_0^4   u_0^{(13)}
\!\! +\! A_0^6   u_0^{(23)}
\right)\rho^2 \, , \quad  \quad 
\nonumber
\\
4 L_2 &\equiv& 
2 (A_0^3)^2  (r_1^M)^2
+\frac{1}{2} (A_0^4)^2  (r_1^M)^2
+\frac{1}{2} (A_0^6)^2  (r_1^M)^2
-A_0^1 A_0^4  r_1^M r_4^M
-3A_0^3 A_0^6  r_1^M r_4^M
\nonumber\\
&& 
-\sqrt{3} A_0^6 A_0^8  r_1^M r_4^M
\!+\frac{1}{2} (A_0^1)^2  (r_4^M)^2
\!+\frac{1}{2} (A_0^3)^2  (r_4^M)^2
\!+\frac{1}{2} (A_0^6)^2  (r_4^M)^2
\!+\sqrt{3} A_0^3 A_0^8 (r_4^M)^2
\nonumber\\
&& 
+\frac{3}{2} (A_0^8)^2  (r_4^M)^2
+3 A_0^3 A_0^4  r_1^M r_6^M
-A_0^1 A_0^6 r_1^M r_6^M
-\sqrt{3} A_0^4 A_0^8 r_1^M r_6^M
-A_0^4 A_0^6  r_4^M r_6^M
\nonumber\\
&& 
+ 2 \sqrt{3} A_0^1 A_0^8 r_4^M r_6^M
+\frac{1}{2} (A_0^1)^2  (r_6^M)^2
+\frac{1}{2} (A_0^3)^2  (r_6^M)^2
+\frac{1}{2} (A_0^4)^2  (r_6^M)^2
\nonumber\\
&& 
-\sqrt{3} A_0^3 A_0^8(r_6^M)^2
+\frac{3}{2} (A_0^8)^2  (r_6^M)^2 \, .
\end{eqnarray}

For getting this expression of $L_{A_0}=L_1 + L_2$, 
we have used the equations
\begin{eqnarray}
r_3^M r_1^M=0\, , \quad
 (r_3^M + \sqrt{3}r_8^M)r_4^M=0\, , \quad
 (r_3^M - \sqrt{3}r_8^M)r_6^M=0\, ,
\end{eqnarray}
to eliminate cross terms between $r_{3,8}$ and $y$ in the Lagrangian.
These can be shown explicitly using the solution of the ADHM constraint 
(\ref{ADHM}) in the expansion of the small $\rho^2/r^2$. The expansion was
studied in detail in \cite{Christ:1978jy}. Using the expression given in
Eq.~(5.13) of 
\cite{Christ:1978jy}, 
it is easy to show the equations above. This elimination of
the cross terms is important for a simplification of the computations.
In fact, we can show later that $L_2$ is not necessary, 
when integrating out $A_0$.

Next we evaluate the contribution from the mass term $(X^4)^2$. 
The alignment (\ref{rthree}) simplifies the ADHM data (\ref{cd1}),
(\ref{cd2}) and (\ref{cd3}) quite a lot. In fact, we have
\begin{eqnarray}
 r_1^M \sigma_M &=& \frac{1}{|r|^2}\rho_1 \rho_2 \; r T_{21}
-\frac{1}{|r|^4}\rho_1\rho_2 \rho_3^2 \; r (T_{32}T_{13}-T_{13}T_{32})
+{\cal O}(1/|r|^5)\, , \\
 r_4^M \sigma_M &=& \frac{2}{|r|^2}\rho_1 \rho_3 \; r T_{31}
-\frac{1}{|r|^4}\rho_1\rho_3 \rho_2^2 \; r (T_{32}T_{12}-T_{12}T_{32})
+{\cal O}(1/|r|^5)\, , \\
 r_6^M \sigma_M &=& -\frac{2}{|r|^2}\rho_2 \rho_3 \; r T_{32}
+\frac{1}{|r|^4}\rho_2\rho_3 \rho_1^2 \; r (T_{31}T_{21}-T_{21}T_{31})
+{\cal O}(1/|r|^5)\, ,
\end{eqnarray}
where $r \equiv r^M \sigma_M$, and 
$T_{ij}\equiv (U^i)^\dagger U^j - (U^j)^\dagger U^i = - T_{ji}$. Note that the
first term in each of the right hand side equals the off-diagonal entry
of the two-body case, $Y$ in (\ref{defY}). The second terms are
corrections due to the three-body effect. So, the three-body
contribution in the mass term ${\rm tr} (X^4)^2$  (\ref{massx4exp})
should appear at the
leading order as a liner term in these second terms, multiplied by the
first terms. An explicit computation leads to
\begin{eqnarray}
&& V^{\rm mass}_{\rm 3-body}
= \frac{\lambda N_c M_{\rm KK}^3}{2^2 3^4 \pi} \frac{\rho^6}{|r|^6}
\nonumber \\
&& \qquad \times 
\bigl(
{\rm tr}[r T_{21}] {\rm tr}[r (T_{23}T_{13}-T_{13}T_{23})]
-2 {\rm tr}[r T_{31}] {\rm tr}[r (T_{32}T_{12}-T_{12}T_{32})]
\bigr.
\nonumber \\
&& \qquad \quad 
\bigl.
-2 {\rm tr}[r T_{32}] {\rm tr}[r (T_{31}T_{21}-T_{21}T_{31})]
\bigr)\, .
\label{3bodysimple2}
\end{eqnarray}
We have already subtracted the one-body and the two-body potentials
here, and took $\rho_1=\rho_2=\rho_3=\rho$ which is the classical value
(the leading value in the large $N_c$ expansion).

\vspace{3mm}
\noindent
\underline{Procedure (3): integrating out $A_0$}
\vspace{3mm}

Once we solve the simultaneous equations of motion for (\ref{simleqs})
for $A_0^{\zeta}$, and plug the solutions into the Lagrangian 
$L_{A_0}$, we should obtain 
\begin{eqnarray}
L_{A_0} = - V\, ,
\quad V \equiv   \sum_{A=1,2,3} V_{\rm 1-body}^{(A)} +
\frac12 \sum_{A\neq B}V_{\rm 2-body}^{(A,B)} + V_{\rm 3-body}
\label{expand3bod}
\end{eqnarray}
where the first term is the one-body rest energy, and the 
second term is the two-body interaction potential. As obtained in
\cite{Hashimoto:2010je}, their expressions are
\begin{eqnarray}
 V_{\rm 1-body}^{(A)} 
= \frac{27\pi N_c}{4\lambda M_{\rm KK}}
\frac{1}{\rho^2} \, , \quad
 V_{\rm 2-body}^{(A,B)} 
= \frac{27\pi N_c}{\lambda M_{\rm KK}}
\frac{(u_0^{(AB)})^2}{|r^{(AB)}|^2 + 2\rho^2 - 2 (u_0^{(AB)})^2 \rho^2}
\, .
\label{12-body}
\end{eqnarray}
Here the inter-nucleon distance is, according to our alignment 
(\ref{rthree}), 
\begin{eqnarray}
 |r^{(12)}|= r \, , \quad  |r^{(13)}|=  |r^{(23)}|= r/2 \, .
\end{eqnarray}
The third term $V_{\rm 3-body}$ is what we like to compute.

We are interested in the regime of short distances $r \ll 1/M_{\rm KK}$.
However, as the classical size of the baryon $\rho$ is quite small, 
$\rho \sim 1/(\sqrt{\lambda} M_{\rm KK})$, the region of our interest is 
rather a ``long-distance'' expansion $\rho \ll r$ in effect. Therefore
we need to expand the resultant Hamiltonian for small $\rho/r$.
As we look at the Lagrangian $L_{A_0}=L_1+ L_2$, we notice that $L_2$
is of order $\rho^4/r^2$, as we know that the ADHM constraint is solved 
in this expansion as $y = {\cal O}(\rho^2/r)$. On the other hand,
As is obvious from (\ref{expand3bod}), the two-body interaction is 
${\cal O}(1/r^2)$ so the three-body interaction should start from
$\rho^2/r^4$. (This is suggested also from the soliton approach, 
see \cite{Hashimoto:2009as}.) 
Therefore, $L_2$ is not necessary as it is at higher order
in this expansion.\footnote{As a check, we can perform 
a computation with keeping $L_2$ explicitly to confirm this. The
computation is lengthy and is not presented in this manuscript, but we
have confirmed it. Note that for the generic case with nonzero $r_8$,
this simplification is not expected, because in general there are terms
of the form $r_8 y$ which contributes additionally to $L_1$, so one
needs explicit expression for $y$ by solving the ADHM constraints. }

The Lagrangian $L_1$ can be conveniently written as
\begin{eqnarray}
 L_1 = \vec{A}^T M \vec{A} + \vec{B}^T \vec{A}
\end{eqnarray}
where 
\begin{eqnarray}
&&
 \vec{A}^T \equiv (A_0^0, A_0^1,A_0^3,A_0^4, A_0^6, A_0^8) \, ,
\quad
\vec{B}^T \equiv \frac{162\pi}{\lambda M_{\rm KK}} (1,0,0,0,0,0)\, ,
\\
&&
M \equiv P + Q, \quad P \equiv 
\frac{r^2}{8} {\rm diag} \; (0,4,0,1,1,0) \, ,
\end{eqnarray}
\begin{eqnarray}
Q \equiv \rho^2
\left(
\begin{array}{cccccc}
6 &  2u_0^{(12)} & 0& 2u_0^{(13)}& 2u_0^{(23)} & 0 \\
2u_0^{(12)} & 1 & 0&u_0^{(23)}/2 &u_0^{(13)}/2 & u_0^{(12)}/\sqrt{3} \\
0 &0 & 1 & u_0^{(13)}/2&-u_0^{(23)}/2 &  0 \\
2u_0^{(13)} & u_0^{(23)}/2&u_0^{(13)}/2 & 1 &u_0^{(12)}/2 & -u_0^{(13)}/2\sqrt{3} \\
2u_0^{(23)} & u_0^{(13)}/2&-u_0^{(23)}/2 & u_0^{(12)}/2& 1 & -u_0^{(23)}/2\sqrt{3} \\
0 &u_0^{(12)}/\sqrt{3}  & 0&-u_0^{(13)}/2\sqrt{3}&-u_0^{(23)}/2\sqrt{3}& 1  
\end{array}
\right)\, . \quad
\end{eqnarray}
Since $M$ is a symmetric matrix {\it i.e.} $M^T = M$, the equations of motion for $A_0$ is solved by
\begin{eqnarray}
 \vec{A} = -\frac12 M^{-1} \vec{B}\, ,
\end{eqnarray}
which is substituted back to $L_1$ to give the Hamiltonian (which is 
$-L_{A_0}$)
\begin{eqnarray}
 V = \frac{\lambda M_{\rm KK} N_c}{54 \pi} \cdot 
\frac{1}{4} \vec{B}^T M^{-1} \vec{B}\, .
\end{eqnarray}
As $\vec{B}$ has only one non-zero entry, this is nothing but
\begin{eqnarray}
 V = 
\frac{3^5\pi N_c}{2\lambda M_{\rm KK}}
\left[M^{-1}\right]_{(1,1)}
\end{eqnarray}
which can be evaluated using the first cofactor of the matrix
$M$. By expanding in power series of $\rho^2/r^2$ up to $O(\rho^4/r^6)$, we obtain, 
\begin{eqnarray}
V = \frac{3^5\pi N_c}{2\lambda M_{\rm KK}} 
\left(  \frac{1}{6 \rho^2} + \frac{2 (u^{(1,2)})^2 + 8 (u^{(1,3)})^2 + 8
 (u^{(2,3)})^2}{9 r^2}  + \frac{4 \rho^2 f_{\rm SI}}{9 r^4} \right)
+ {\cal O} \left(\rho^4/{r^6}\right) , \, \, \quad \,
\end{eqnarray}
where spin/isospin phase $f_{\rm SI}$ is defined as 
\begin{eqnarray}
f_{\rm SI} 
&\equiv
& (u^{(1,2)}_0)^4  - (u^{(1,2)}_0)^2 + 16 (u^{(1,3)}_0)^4 - 16 (u^{(1,3)}_0)^2 + 16 (u^{(2,3)}_0)^4 -16 (u^{(2,3)}_0)^2 
\nonumber \\
&& 
+ 4  (u^{(1,2)}_0)^2  (u^{(1,3)}_0)^2 + 4 (u^{(1,2)}_0)^2 (u^{(2,2)}_0)^2 + 16 (u^{(1,3)}_0)^2 (u^{(2,3)}_0)^2
 \nonumber \\
&& - 24 u^{(1,2)}_0 u^{(2,3)}_0u^{(1,3)}_0 \,.
\end{eqnarray}
Subtracting the 1-body and 2-body potentials 
(\ref{12-body}) from this expression as in (\ref{expand3bod}), we obtain 
the potential intrinsic to the three-body nature by the expansion of $\rho^2/r^2$, 
which we call $V_{\rm 3-body}^{A_0}$ as   
\begin{eqnarray}
V_{\rm 3-body}^{A_0} &=&  
\frac{216 \pi N_c\rho^2}{\lambda M_{\rm KK}|r|^4}
\left[
(u^{(1,2)}_0)^2 (u^{(1,3)}_0)^2
+ (u^{(1,2)}_0)^2 (u^{(2,3)}_0)^2
+ 4 (u^{(1,3)}_0)^2 (u^{(2,3)}_0)^2
\right.
\nonumber \\
&& 
\left.\hspace{30mm}
-6  u^{(1,2)}_0 u^{(2,3)}_0u^{(1,3)}_0
\right] + {\cal O}(\rho^4/r^6) \, .
\label{3bodysimple}
\end{eqnarray}

With the potential coming from the $X^4$ mass term (\ref{3bodysimple2}), 
the total three-body potential is 
\begin{eqnarray}
 V_{\rm 3-body} = V_{\rm 3-body}^{A_0}+ V_{\rm 3-body}^{\rm mass}\, .
\end{eqnarray}
With this at hand, we can evaluate this potential with any three-baryon
state with any spin/isospin. Next, we shall choose 
two wave functions, one is appropriate for the neutron stars, 
and the other is for a Helium-3 nucleus and a triton (a nucleus of tritium), 
to find the three-body nuclear
potential is positive. 

We have two remarks on (\ref{3bodysimple}). First, the three-body
Hamiltonian (\ref{3bodysimple}) is of order ${\cal O}(1/(\lambda^2 r^4))$
because of 
$\rho\sim 1/\sqrt{\lambda}$, and so it is suppressed by $1/\lambda^2$. 
This is consistent with the generic
observation given in the soliton picture \cite{Hashimoto:2009as} 
stating that the generic $k$-body potential is of order 
$1/(\lambda^{k-1}r^{2k-2})$ in the unit of $M_{\rm KK}=1$.
Second, in the expression above if we put all
the matrices $U^{(i)}$ equal to each other so that the ADHM data is that
of the 'tHooft instantons, we have $u_0^{(i,j)}=1$ 
and $A_{ij}=0$, resulting in 
the vanishing three-body potential. This 
is consistent with the result of
the previous section.

\vspace{3mm}
\noindent
\underline{Procedure (4): evaluate the Hamiltonian with
baryon states}
\vspace{3mm}

Now, we are ready to compute the spin/isospin dependence of the
three-body nuclear force at short distances. Although we can evaluate it
for any choice of spin/isospin for each baryon, in this paper 
we choose
the following two states as explicit examples: 
\begin{itemize}
 \item[(4-a)] three neutrons with spins averaged.
\item[(4-b)] proton-proton-neutron (and proton-neutron-neutron). 
 \end{itemize}
The reason for these choices is that 
the first example is relevant for dense states of 
many neutrons such as core of neutron stars and 
supernovae, where the three-body nuclear forces are quite important. 
The second is obviously for the spectrum of Helium-3 nucleus where 
three-body forces are expected to contribute, and also for a triton. 

\vspace{3mm}
\noindent
\underline{(4-a): three neutrons with spins averaged}
\vspace{3mm}

For protons and neutrons, the single-baryon wave function 
is given by (\ref{wave}). 
For the neutron stars and the supernovae, 
we need neutron states with spins averaged. Thus, for any given operator
of the quantum mechanics, the appropriate expectation value for these 
is obtained by 
\begin{eqnarray}
 \langle V \rangle = 
\frac12 
\left[
\langle n\uparrow | \widehat{\cal O}|n\uparrow\rangle
+
\langle n\downarrow | \widehat{\cal O}|n\downarrow\rangle
\right] 
\label{singleO}
\end{eqnarray}
For the case of $\widehat{\cal O}$ being the three-body Hamiltonian,
we need to take the above expectation value for each of three baryons. 
As nucleons are fermions, any wave functions should be anti-symmetric
under the exchange of the nucleons. 
Here, as three neutrons move around in realistic situations, we do not
anti-symmetrize the wave functions\footnote{In fact, once
we take three neutrons for the isospin sector, it is impossible to
anti-symmetrize the wave function with the spin sector, without a help
of angular momenta.} (in this paper we have not evaluated
nuclear potentials coming from motion of the baryons).

Here for a demonstration, let us consider a single baryon case
(\ref{singleO}). 
Using the coordinate expression of the wave functions, this
(\ref{singleO}) means
\begin{eqnarray}
 \langle V \rangle 
= \int d\Omega_3 \;
\frac12 
\biggl[{\cal O} \;
|\langle \vec{a}|n\uparrow\rangle|^2
+
 {\cal O} \; 
|\langle \vec{a}|n\downarrow\rangle|^2
\biggr] 
\end{eqnarray}
Here $d\Omega_3$ is the integration over the $S^3$ spanned by the unit 
vector $\vec{a}$. Using the wave functions (\ref{wave}), we can see
\begin{eqnarray}
|\langle \vec{a}|n\uparrow\rangle|^2
+|\langle \vec{a}|n\downarrow\rangle|^2
= \frac{1}{\pi^2} \Bigl[
 (a_1)^2 + (a_2)^2 + (a_3)^2 + (a_4)^2 
\Bigr]
= \frac{1}{\pi^2}
\end{eqnarray}
So, we obtain a simple expression
\begin{eqnarray}
 \langle V \rangle 
= \frac{1}{2\pi^2}\int \! d\Omega_3 \; {\cal O}.
\end{eqnarray}

Using this simple formula, the three-body potential with the 
spin-averaged wave function is
\begin{eqnarray}
\lefteqn{\left\langle
 V_{\rm 3-body}^{A_0}\right\rangle_{\rm nnn(spin-averaged)}}
\nonumber \\
 &&=  
\frac{216 \pi N_c\rho^2}{\lambda M_{\rm KK}|r|^4}
\frac{1}{(2\pi^2)^3}
\int \! d\Omega_3^{(1)}d\Omega_3^{(2)}d\Omega_3^{(3)}
\left[
(u^{(1,2)}_0)^2 (u^{(1,3)}_0)^2
+ (u^{(1,2)}_0)^2 (u^{(2,3)}_0)^2
\right.
\nonumber
\\
&&\hspace{50mm} \left.
+ 4 (u^{(1,3)}_0)^2 (u^{(2,3)}_0)^2
-6  u^{(1,2)}_0 u^{(2,3)}_0u^{(1,3)}_0
\right].
\end{eqnarray}
This integral over three $S^3$'s can be easily performed. 
For example, 
for $(u_0^{(1,2)})^2$, using the definition below (\ref{wave}) and
(\ref{2bodya}), we get  
\begin{eqnarray}
u_0^{(i,j)} 
= \frac{1}{2} {\rm tr}\left[U^{(i)\dagger}U^{(j)}\right] = \vec a^{(i)}
\cdot \vec a^{(j)} \,, 
\end{eqnarray}
where $\vec a^{(i)}$ is unit 4-component vector, pointing one phase point on $S^3$ for spin/isospin d.o.f. $U^{(i)}$.   
Therefore, we obtain
\begin{eqnarray}
\int \! d\Omega_3^{(1)}  (u_0^{(1,2)})^2 
= \int \! d\Omega_3^{(1)}  \cos^2\theta
= \int \! \cos^2\theta
\sin^2\theta \sin\tilde{\theta} d\theta d\tilde{\theta} 
d\tilde{\tilde{\theta}} = \frac{\pi^2}{2}\, ,
\end{eqnarray}
where $\theta$ is the angle between $\vec{a}^{(1)}$ and 
$\vec{a}^{(2)}$. Using this and also the following integral
\begin{eqnarray}
\int \! d\Omega_3^{(1)}d\Omega_3^{(2)}d\Omega_3^{(3)}
 u^{(1,2)}_0 u^{(2,3)}_0u^{(1,3)}_0 = \frac{\pi^6}{2} \, , 
\end{eqnarray}
we obtain 
\begin{eqnarray}
\left\langle
V_{\rm 3-body}^{A_0}
\right\rangle_{\rm nnn(spin-averaged)} = 0 
\, .
\end{eqnarray}
Therefore the three-body potential from the $A_0$ term vanishes for the 
spin-averaged neutron wave function.

In the same manner, for $V_{\rm 3-body}^{\rm mass}$, the expectation
value is given as 
\begin{eqnarray}
 && 
\left\langle
V_{\rm 3-body}^{\rm mass}
\right\rangle_{\rm nnn(spin-averaged)} = 
\frac{\lambda N_c M_{\rm KK}^3}{2^2 3^4 \pi} \frac{\rho^6}{|r|^6}
\nonumber \\
&& \qquad \times\frac{1}{(2\pi^2)^3}\int \!
d\Omega_3^{(1)} d\Omega_3^{(2)} d\Omega_3^{(3)} 
\Bigl(
{\rm tr}[r T_{21}] {\rm tr}[r (T_{23}T_{13}-T_{13}T_{23})]
\Bigr.
\nonumber \\
&& \qquad \quad 
\Bigl.
-2 {\rm tr}[r T_{31}] {\rm tr}[r (T_{32}T_{12}-T_{12}T_{32})]
-2 {\rm tr}[r T_{32}] {\rm tr}[r (T_{31}T_{21}-T_{21}T_{31})]
\Bigr)\, 
\nonumber \\
&&
= -\frac{\lambda N_c M_{\rm KK}^3}{2^2 3^3 \pi(2\pi^2)^3} 
\frac{\rho^6}{|r|^6}
\int \!
d\Omega_3^{(1)} d\Omega_3^{(2)} d\Omega_3^{(3)} 
\Bigl(
{\rm tr}[r T_{21}] {\rm tr}[r (T_{23}T_{13}-T_{13}T_{23})]
\Bigr) \, .\qquad
\end{eqnarray}
Here in the last equality we have used the invariance under the exchange
of the integration variables, 
$d\Omega_3^{(1)} \leftrightarrow d\Omega_3^{(2)} \leftrightarrow 
d\Omega_3^{(3)}$.
The integration can be performed by 
using the polar coordinates of the $S^3$, and the result is
\begin{eqnarray}
 \frac{1}{(2\pi^2)^3}\int \!
d\Omega_3^{(1)} d\Omega_3^{(2)} d\Omega_3^{(3)} 
\Bigl(
{\rm tr}[r T_{21}] {\rm tr}[r (T_{23}T_{13}-T_{13}T_{23})]
\Bigr) = -8 |\vec{r}|^2 \, ,
\end{eqnarray}
where $\vec{r}=(r^1, r^2, r^3)$ is the three-dimensional vector which
specifies the inter-baryon distance in our space. 
At the leading order in $1/N$ expansion, 
we may use the classical value for $r^4$ which is zero, so
in effect the three-dimensional distance is equal to the
four-dimensional one,  $|\vec{r}| = |r|$.
We can substituting the classical value 
$\rho = 2^{-1/4}3^{7/4}\sqrt{\pi}\lambda^{-1/2} M_{\rm KK}^{-1}$ 
at the leading order in the $1/N_c$ expansion.
So, we obtain the three-body potential due to the matrix model 
mass term as 
\begin{eqnarray}
\left\langle
V_{\rm 3-body}^{\rm mass}
\right\rangle_{\rm nnn(spin-averaged)} = 
\frac{2^{-1/2}3^{15/2}\pi^2 N_c}{\lambda^2 M_{\rm KK}^3|r|^4}
\, .
\label{final2}
\end{eqnarray}

Therefore, in total, we obtain  
\begin{eqnarray}
\left\langle
V_{\rm 3-body}
\right\rangle_{\rm nnn(spin-averaged)} &=& 
\left\langle
V_{\rm 3-body}^{A_0}
\right\rangle_{\rm nnn(spin-averaged)} 
+\left\langle
V_{\rm 3-body}^{\rm mass}
\right\rangle_{\rm nnn(spin-averaged)} \nonumber \\
&=& 
\frac{2^{-1/2}3^{15/2} \pi^2 N_c }{\lambda^2 M_{\rm KK}^3|r|^4}
\, .
\label{final3}
\end{eqnarray}
This is the three-body nuclear 
potential for three neutrons placed on a line with equal
spacings $|r|/2$, with spins averaged. 
The three-body potential is
suppressed compared to the two-body potential by $\sim 1/\lambda (r M_{KK})^2 \ll 1$ for 
large $\lambda$, which is generic hierarchy between $N+1$-body potential to $N$-body one 
as shown in \cite{Hashimoto:2009as}. $M_{\rm KK}$ roughly indicates the QCD
scale, and our computation is
valid at short-distance, 
$1/(\sqrt{\lambda}M_{\rm KK}) \ll |r| \ll 1/M_{\rm KK}$.\footnote{This $M_{\rm KK}$ is about 1
GeV if it is fit with $\rho$ meson mass \cite{Sakai:2004cn}, 
while it is about 0.5 GeV when it is fit with baryon mass differences
\cite{Hata:2007mb}. We are working in the large 
$\lambda$ expansion. The 'tHooft coupling constant of QCD,
$\lambda$, is ${\cal O}(10-20)$ when it is 
fit with pion decay constant \cite{Sakai:2004cn}.}

\vspace{3mm}
\noindent
\underline{(4-b): proton-proton-neutron}
\vspace{3mm}

Let us evaluate the three-body potential with the case of
proton-proton-neutron.
We are interested in the three-nucleon state with
a total spin $1/2$ and a total isospin $1/2$. 
For any choice of the third component of the spin/isospins,
we can find a unique wave function with 
a complete anti-symmetrization. 

The proton-proton-nuetron means the third component of the total isospin
is $+1/2$. 
For example, when the third component of the total spin is $+1/2$,
\begin{eqnarray}
&& \frac{1}{\sqrt{6}}
\Bigl[
|p\uparrow\rangle_1 |p\downarrow\rangle_2 |n\uparrow\rangle_3
-
|p\downarrow\rangle_1 |p\uparrow\rangle_2 |n\uparrow\rangle_3
- 
|p\uparrow\rangle_1 |n\uparrow\rangle_2 |p\downarrow\rangle_3
\Bigr.
\nonumber 
\\
&&
\Bigl.
\qquad 
+
|p\downarrow\rangle_1 |n\uparrow\rangle_2 |p\uparrow\rangle_3
-
|n\uparrow\rangle_1 |p\downarrow\rangle_2 |p\uparrow\rangle_3
+
|n\uparrow\rangle_1 |p\uparrow\rangle_2 |p\downarrow\rangle_3
\Bigr] \, .
\end{eqnarray}
The calculation with this wave function is straightforward,
and we find the integrals
\begin{eqnarray}
&&
\int \! d\Omega_3^{(1)}d\Omega_3^{(2)}d\Omega_3^{(3)} 
|\psi(\vec{a}_1, \vec{a}_2, \vec{a}_3)|^2
(u^{(1,2)}_0)^2 (u^{(1,3)}_0)^2
= \frac{1}{36} \, ,
\nonumber
\\
&&
\int \! d\Omega_3^{(1)}d\Omega_3^{(2)}d\Omega_3^{(3)} 
|\psi(\vec{a}_1, \vec{a}_2, \vec{a}_3)|^2
u^{(1,2)}_0 u^{(2,3)}_0u^{(1,3)}_0
= \frac{1}{36}\, ,
\\
&& \int \!
d\Omega_3^{(1)} d\Omega_3^{(2)} d\Omega_3^{(3)} 
|\psi(\vec{a}_1, \vec{a}_2, \vec{a}_3)|^2
\Bigl(
{\rm tr}[r T_{21}] {\rm tr}[r (T_{23}T_{13}-T_{13}T_{23})]
\Bigr) = -\frac{320}{27} |\vec{r}|^2 \, . \quad
\nonumber
\end{eqnarray}
Using these formula, we obtain again
\begin{eqnarray}
\left\langle
V_{\rm 3-body}^{A_0}
\right\rangle_{\rm ppn} = 0 
\, ,
\end{eqnarray}
while for the other potential we have a different factor
\begin{eqnarray}
\left\langle
V_{\rm 3-body}^{\rm mass}
\right\rangle_{\rm ppn} = 
\frac{2^{5/2}3^{9/2}5\pi^{2} N_c}{\lambda^2 M_{\rm KK}^3|r|^4}
\, .
\label{final2ppn}
\end{eqnarray}
Therefore, in total, the three-body potential is 
\begin{eqnarray}
\left\langle
V_{\rm 3-body}
\right\rangle_{\rm ppn} &=& 
\left\langle
V_{\rm 3-body}^{A_0}
\right\rangle_{\rm ppn} 
+\left\langle
V_{\rm 3-body}^{\rm mass}
\right\rangle_{\rm ppn} =
\frac{2^{5/2}3^{9/2}5 \pi^{2} N_c }{\lambda^2 M_{\rm KK}^3|r|^4}
\, .
\label{final3ppn}
\end{eqnarray}
The three-body potential is positive, that means, we have a repulsive
three-body force at short distances.

This computation is for $(+1/2, +1/2)$ of the third components of the
spin and the isospin. Computations with three 
other wave functions, $(+1/2, -1/2)$, $(-1/2,+1/2)$, and $(-1/2,-1/2)$,
can be done in the same manner, and the result for the three-body
potential turns out to be the same as (\ref{final3ppn}) 
for all of these. These are due to the fact that the action (\ref{mm}) has 
rotational invariance $SO(3)$ and isospin $SU(2)$ invariance. 
This includes the case
for proton-neutron-neutron, which is the case for a triton  
(a tritium nucleus).


\vspace{2mm}
\section{Summary and discussions.} 
\label{sec4}

With the simple $U(k)$ matrix model for $k$-nucleon systems which we 
proposed in \cite{Hashimoto:2010je} together with P.~Yi, 
in this paper we have computed 
short-distance three-body nuclear forces. Our matrix model is  
not a phenomenological model, but 
derived in string theory using the gauge/string duality (the AdS/CFT
correspondence). 
More precisely, our matrix model is a low-energy
effective field theory on baryon vertex D4-branes 
\cite{Witten:1998xy} in the D4-D8
holographic model \cite{Sakai:2004cn,Sakai:2005yt} of large $N_c$ QCD. 
In this framework, we can compute nuclear potentials for arbitrary number 
$k$ of the nucleons.

Our computations are straightforward. For three nucleons, we took
$k=3$, {\it i.e.}~$U(3)$ matrix model. The matrix model 
Hamiltonian evaluated with a quantum three-baryon state, 
a tensor product of single-baryon states, gives the three-body nuclear
potential. We subtracted one-body and two-body contributions, thus the
remaining is the force intrinsic to the three-body. The computations are
valid only at short range, $1/\sqrt{\lambda} M_{\rm KK} \ll |r| \ll
1/M_{\rm KK}$ where $\lambda$ is the 'tHooft coupling constant of the 
QCD which is ${\cal O}(10)$ for fitting pion decay constant
\cite{Sakai:2004cn}, and $M_{\rm KK}$ is ${\cal O}(1-0.5)$ GeV  
\cite{Sakai:2004cn,Hata:2007mb} when it is fit with meson/baryon
masses (or mass differences). 
As explicit 
examples, we took  a) 
three neutrons with spins averaged, and b)  
proton-proton-neutron, both aligned on a line with equal
spacings. 
The resultant three-body nuclear
potentials are (\ref{final3})
and (\ref{final3ppn}), both of which are 
positive. 

Let us discuss possible importance of our result. 
We have computed (\ref{final3}) for three-body neutrons. But as seen from 
the form of wave functions (\ref{wave}) and isospin $SU(2)$ invariance of the action (\ref{mm}), 
the results hold also for three-body forces for three-protons. Therefore the result (\ref{final3}) hold as 
far as all three nucleons have same flavor. 
In the same manner, the
three-body potential (\ref{final3ppn}) for the proton-proton-neutron 
is equal to the three-body potential for 
proton-neutron-neutron, which is 
responsible for a triton. These results imply that there are additional 
repulsive forces in addition to two-body forces for these states at short distances.  

The three-body potentials which we obtained in (\ref{final3}) 
and (\ref{final3ppn}) are suppressed by $1/\lambda (r M_{KK})^2$ 
compared with two-body potential, and at the length scale where our computation 
is valid, {\it i.e.} $1/(\sqrt{\lambda}M_{\rm KK}) \ll |r| \ll 1/M_{\rm KK}$, this suppression 
factor $1/\lambda (r M_{KK})^2$ is small. 
This makes our three-body potential computation 
valid; we have two-body dominant repulsive potential and furthermore 
small but nonzero repulsive potential from three-body forces. 

These three-body forces are stronger as distances get  
shorter. As a result, at very short distances where neutrons are 
highly dense, three-body forces give additional 
repulsive forces. 
This statement supports recent observation that the nuclear
two-body repulsion is not enough to explain supernovae explosions, nor 
the equations of states for the core of neutron stars.
In high density nuclear matters such as the neutron stars, 
our result suggests 
that the repulsive core of neutrons in neutron stars and supernovae 
has an extra positive contribution besides the repulsive potential from 
the two-body nuclear potential. 
The necessity of the repulsive three-body forces for neutrons
has been indicated by analysis of mass bounds 
of neutron stars and supernova explosion simulations. 

We also found that the three-body forces for proton-proton-neutron at
short distances 
is repulsive. 
In Helium-3 nuclear spectrum, it is expected that 
a short-range repulsive three-body forces is necessary, and our result
sounds to be consistent with this. Furthermore, we found a repulsive
three-body forces also for a proton-neutron-neutron, which should be
related to a triton.
There are other related issues 
in few-body nuclear spectra\footnote{
Spectra of heavy nuclei have been discussed in a holographic approach 
in \cite{Hashimoto:2008jq}.}.

Our example is limited to three nucleons on a line, 
so this is not conclusive for the
questions concerning the interesting situations listed above. 
However, our results are 
suggestive. In principle, it is very straightforward to 
compute the $k$-body forces at arbitrary arrangements of nucleons using our matrix model,  
therefore this matrix model is effective for studying short-range many-body
nuclear forces.


\acknowledgments
We would like to thank P.~Yi for collaboration 
on the earlier project \cite{Hashimoto:2010je}. 
We would also like to thank S.~Aoki, 
T.~Doi, T.~Hatsuda, T.~Nakatsukasa, K.~Sekiguchi, and G.~Watanabe 
for helpful
discussions and comments. 
K.H.~and N.I.~would like to thank Yukawa Institute for Theoretical Physics.  
K.H.~thanks CERN. N.I.~also thanks RIKEN. 
This research was partially supported by 
KAKENHI Grant-in-Aid 21105514, 19740125, 22340069.

\end{document}